\documentclass[USenglish]{article}	
\usepackage[utf8]{inputenc}				%(only for the pdftex engine)
\usepackage[big,online]{dgruyter}	%values: small,big | online,print,work
\usepackage{lmodern} 
\usepackage{microtype}

\usepackage[numbers,square,sort&compress]{natbib}
\usepackage{gensymb}
\begin{document} 
%tomyself-font in the figures is 20 Arial
%%%--------------------------------------------%%%
	\articletype{Research Article}
%	\received{Month	DD, YYYY}
%	\revised{Month	DD, YYYY}
%  \accepted{Month	DD, YYYY}
%  \journalname{De~Gruyter~Journal}
%  \journalyear{YYYY}
%  \journalvolume{XX}
%  \journalissue{X}
%  \startpage{1}
%  \aop
%  \DOI{10.1515/sample-YYYY-XXXX}
%%%--------------------------------------------%%%

\title{Direct Growth of Monolayer MoS$_{2}$ on Nanostructured Silicon Waveguides}
%\runningtitle{Short title}
%\subtitle{Insert subtitle if needed}
%\ use * to mark the author as the corresponding author
\author*[1]{Athira Kuppadakkath}
\author*[2]{Emad Najafidehaghani}
\author[3]{Ziyang Gan} 
\author[3]{Alessandro Tuniz} 
\author[3]{Gia Quyet Ngo} 
\author[3]{Heiko Knopf} 
\author[3]{Franz J. F. Löchner} 
\author[3]{Fatemeh Abtahi} 
\author[3]{Tobias Bucher} 
\author[3]{Sai Shradha} 
\author[3]{Thomas Käsebier} 
\author[3]{Stefano Palomba} 
\author[3]{Nadja Felde} 
\author[3]{Pallabi Paul} 
\author[3]{Tobias Ullsperger} 
\author[3]{Sven Schröder} 
\author[3]{Adriana Szeghalmi} 
\author[3]{Thomas Pertsch} 
\author[3]{Isabelle Staude} 
\author[3]{Uwe Zeitner} 
\author[3]{Antony George} 
\author*[2]{Andrey Turchanin} 
\author*[1]{Falk Eilenberger}

%\runningauthor{F.~Author et al.}
\runningauthor{A. Kuppadakkath et al.}
\affil[1]{\protect\raggedright 
Friedrich Schiller University, Institute of Applied Physics, Jena, Germany, e-mail: falk.eilenberger@uni-jena.de}
\affil[2]{\protect\raggedright 
Friedrich Schiller University, Institute of Physical Chemistry, Jena, Germany, e-mail: andrey.turchanin@uni-jena.de}
%\communicated{...}
%\dedication{...}
	
\abstract{We report for the first time the direct growth of Molybdenum disulfide (MoS$_{2}$) monolayers on nanostructured silicon-on-insulator waveguides. Our results indicate the possibility of utilizing the Chemical Vapour Deposition (CVD) on nanostructured photonic devices in a scalable process. Direct growth of 2D material on nanostructures rectifies many drawbacks of the transfer-based approaches. We show that the van der Waals materials grow conformally across the curves, edges, and the silicon-SiO$_{2}$ interface of the waveguide structure. Here, the waveguide structure used as a growth substrate is complex not just in terms of its geometry but also due to the two materials (Si and SiO$_{2}$) involved. A transfer-free method like this yields a novel approach for functionalizing nanostructured, integrated optical architectures with an optically active direct semiconductor.}


\maketitle
\section{Introduction} 
Silicon is an indirect semiconductor with a centrosymmetric crystalline structure, which is not ideal for detection, electro-optic modulation, or light emission \cite{lockwood_silicon_2004,soref_past_2006,tsybeskov_silicon_2009,dong_silicon_2014}. Therefore, CMOS-based optoelectronic chips, which utilize optical waveguides, have remained elusive. It is challenging to integrate the well-established direct semiconductors such as III-V semiconductors due to the epitaxial mismatch with silicon. On the other hand, semiconducting Van der Waals materials can be grown on a large set of non-epitaxial substrates, yielding, e.g., field-effect transistors and other electro-optic devices \cite{zhu_electronic_2014,liu_rotationally_2016,bergeron_chemical_2017,vogl_radiation_2019}. Therefore, direct integration of these 2D materials into silicon nanostructures in a wafer scalable process may add nanoscopic light sources and detectors into the toolbox of silicon photonics, thus helping bridge the gap between the excellent electronic properties of silicon and its established waveguide capabilities. 

Transition metal dichalcogenides (TMDs) have the general formula MX$_{2}$, where M represents the transition metal, and X represents a chalcogen \cite{wilson_transition_1969,chhowalla_chemistry_2013,manzeli_2d_2017}. Most TMDs can be exfoliated into monolayers because individual layers are bonded by the weak van der Waals force \cite{wilson_transition_1969}. TMDs are semiconductors \cite{wilson_transition_1969} with a bandgap in the near-infrared or long-wavelength part of the visible spectrum. Monolayer TMDs are direct bandgap semiconductors \cite{mak_atomically_2010} with remarkable optical and electronic properties. They exhibit strong photoluminescence (PL) \cite{mak_atomically_2010,splendiani_emerging_2010,eda_photoluminescence_2011,gutierrez_extraordinary_2013,amani_near-unity_2015} driven by excitons. The TMDs having a single or odd number of layers are highly nonlinear \cite{kumar_second_2013,malard_observation_2013,li_probing_2013} due to the lack of inversion symmetry. Planar heterojunctions of TMDs are also suitable as light sources and detectors \cite{bie_mote2-based_2017,najafidehaghani_1d_nodate}. 

Different groups have integrated 2D materials on waveguides and other photonic-nanostructures utilizing the transfer process\cite{liu_enhanced_2015,chen_enhanced_2017,vogl_room_2017,zhang_optical-resonance-enhanced_2018,bucher_tailoring_2019,peyskens_integration_2019,zhang_enhanced_2020,lochner_hybrid_2021}. However, the transfer process is hard to reproduce and difficult to scale. It is also inherently limited to the top surface of the nanostructure, which means that functionalizing sidewalls or under-etched geometries is very challenging. Furthermore, transfer-based approaches cannot ensure a homogeneous contact between the 2D material and the waveguide structure without creating any bubbles or wrinkles \cite{calado_formation_2012,uwanno_fully_2015}. Transferring on structured substrates also induces strain fields \cite{rao_spectroscopic_2019}. Although the strain in 2D TMDs is beneficial for creating single-photon emitters \cite{tonndorf_single-photon_2015,kumar_strain-induced_2015,branny_discrete_2016,kern_nanoscale_2016,branny_deterministic_2017,palacios-berraquero_large-scale_2017,carmesin_quantum-dot-like_2019,mukherjee_electric_2020,fuchs_controlling_2021}, it is challenging to reproduce.

Previous works \cite{sun_direct_2019,bae_integration_2019,yu_van_2020} have demonstrated the use of van der Waals epitaxy for the integration of 2D materials with different structures. In that context, the 2D material acts as the starting material. Then bulk Si or GaAs is grown on it by classic epitaxy, albeit not at the quality required for many optoelectronic applications. Nevertheless, this suggests that van der Waals materials can be integrated with these materials because their binding properties facilitate crystalline growth without epitaxially compatible substrates. 

In contrast to van der Waals epitaxy, we fabricated the silicon on insulator (SOI) waveguides in the first step. Then we grew TMD monolayer crystals directly on the nanostructure using Chemical vapor deposition (CVD). CVD is a well-established method to synthesize crystalline TMD monolayers \cite{lee_synthesis_2012,li_centimeter-scale_2018,lan_wafer-scale_2018,george_controlled_2019,ngo_scalable_2020} or even heterostructures \cite{george_giant_2021} with a material quality comparable to that of exfoliated samples \cite{george_controlled_2019}. Since the CVD process of TMDs is based on non-directional chemisorption, the crystals grow conformally over the nanostructure following the bends and corners of the geometry as shown in Figure 1. Both silicon \cite{cullis1985microscopy} and SiO$_{2}$ \cite{zhong20} won't undergo deformation at the temperature and pressure required for the CVD growth of MoS$_{2}$. Therefore this approach is assumed to be compatible with established CMOS processes for silicon nanostructures. Although the growth temperature for hBN is much higher compared to TMDs ($\approx$ 1000 $\degree$C) \cite{rice_effects_2018,jeong_wafer-scale_2019}, recent work demonstrated the direct growth of hBN on photonic chips \cite{glushkov_direct_2021}. However, hBN is a wide-gap dielectric, and electro-optical functionalization is thus difficult \cite{vogl_atomic_2019}. 

\begin{figure}[h!] 
\centering
\includegraphics[width=1\columnwidth]{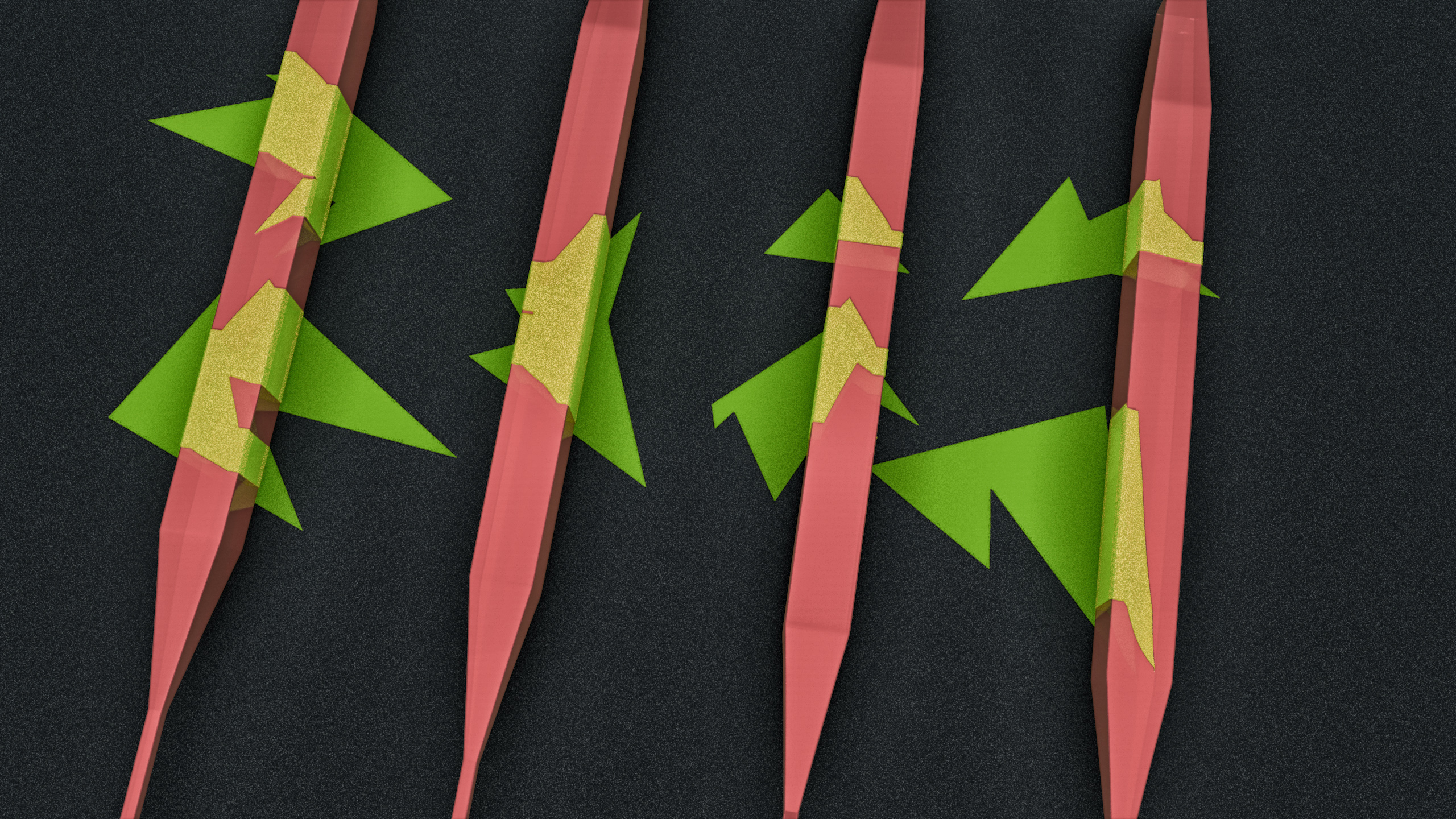}
\caption{Concept schematic of the MoS$_{2}$ crystal (green) grown over a Si waveguide structure (pink) via the CVD process.}
\label{fig1}
\end{figure}

\section{Results and Discussion} 
The nanostructured sample consisted of silicon ridge waveguides tapering to grating couplers at both ends \cite{tuniz_modular_2020}. The waveguides were fabricated with a multistep electron-beam lithography technique on 220 nm silicon-on-insulator substrate, later diced into 10$\times$10 mm chips. The waveguides and grating couplers were designed to operate at a wavelength of 1320 nm. The waveguides were still functional after the high temperature (730 $\degree$C) CVD growth process of the MoS$_{2}$ crystals (Supplementary 4). In this study we resorted to an in-depth analysis of the growth and characterization of the TMDs on the nanostructured substrate.

\begin{figure}[h!] 
\centering
\includegraphics[width=1\columnwidth]{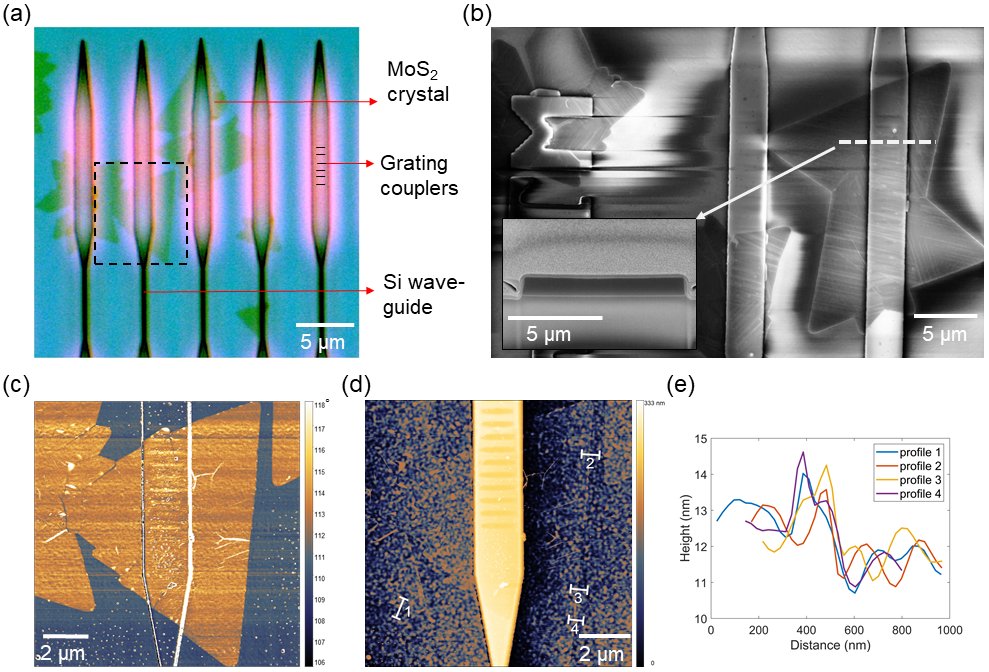}
\caption{Morphological characterization of MoS$_{2}$ crystals on SOI waveguides. a) Optical microscope image of CVD-grown MoS$_{2}$ on waveguides. The blue areas are the SiO$_{2}$ substrate. The pink, wider parts at the top consist of Si grating couplers and the black narrower parts below are the Si waveguides. MoS$_{2}$-crystals appear as greenish. The waveguides are 220 nm higher than the substrate, therefore the top of the waveguides is slightly out of focus. b) Scanning electron microscopic (SEM) image of MoS$_{2}$ crystals grown seamlessly over edges and materials boundaries. (inset) The focused ion beam (FIB) cross section of the waveguide along the dotted line in the SEM image to highlight the nanoscale bending radius of the edges. c) Phase image from atomic force microscopic (AFM) measurement of a region marked by the box in Figure 2a. d) AFM height map corresponding to the AFM phase image. The grains in the height map are due to the surface roughness (RMS value 2 nm) of the substrate. e) AFM height profiles across the lines marked in the height map in Figure 2d.}
\label{fig2}
\end{figure} 

MoS$_{2}$ monolayers were grown on the waveguide chips using a well-established CVD technique \cite{george_controlled_2019}. The TMDs in their characteristic angular shape had lateral dimensions ranging from 2 ${\mu}$m to 10 ${\mu}$m, as seen in the optical microscope image (Figure 2a). The crystals grew on the SiO$_{2}$ substrate, the waveguides, the grating couplers, and even the fabrication labels. In general, we observed more MoS$_{2}$ crystals on the grating couplers (pink regions in Figure 2a) compared to the waveguides themselves (black lines in Figure 2a). We attribute this to the couplers providing more surface roughness due to the inscribed shallow gratings and thus providing more nucleation sites for the initial growth of the TMDs. 

We did not observe any apparent interruption in the individual crystals at a material interface between Si and SiO$_{2}$, edges or corners in the optical microscope image (Figure 2a). The continuity of the crystals is more evident in the scanning electron microscopy (SEM) image in Figure 2b. The crystals wrapped over a “TM” shaped nanostructure (top left of the SEM image), where the termination of the Si-nanostructure is complex. The inset of Figure 2b shows the SEM image of the focused ion beam milled (FIB) cross-section across the grating coupler with MoS$_{2}$ grown on it. This reveals the nanoscale convex bends at the top (radius of curvature <30 nm) and concave bends at the bottom (radius of curvature < 25 nm). The crystal grew over these bends, along the 220 nm high sidewalls, and across the Si-SiO$_{2}$ interface at the bottom. 

To prove the monolayer nature of the as-grown MoS$_{2}$, we performed multiple AFM measurements. The phase image in Figure 2c clearly shows the outline of the crystal. The phase of the crystal is the same on the SiO$_{2}$ substrate as well as on the Si waveguide. This suggests that the properties of the MoS$_{2}$ crystals are unaffected by the growth substrate and not deteriorated by the nanostructure. The height map related to the phase image is shown in Figure 2d. The surface roughness of the waveguide structure is high (RMS= 2 nm) compared to the thickness of monolayer MoS$_{2}$ due to the waveguide fabrication process. Therefore it is difficult to observe the TMD-crystal clearly in topography mode. However, we can see the edges of the crystals clearly after filtering. The height profiles shown in Figure 2e, shows an average thickness of 1 nm for the monolayer crystals. The Raman spectra measurements also confirms the layer number (Supplementary 3).

This shows that monolayer TMD-crystals can be grown on nanostructures defying the epitaxial matching required for common 3D semiconductors. Particularly, the growth around a bend with radius less than 30 nm is well beyond prior observations of CVD growth on fiber core with a bending radius above 10 µm \cite{ngo_scalable_2020}.

\begin{figure}[h!] 
\centering
\includegraphics[width=1\columnwidth]{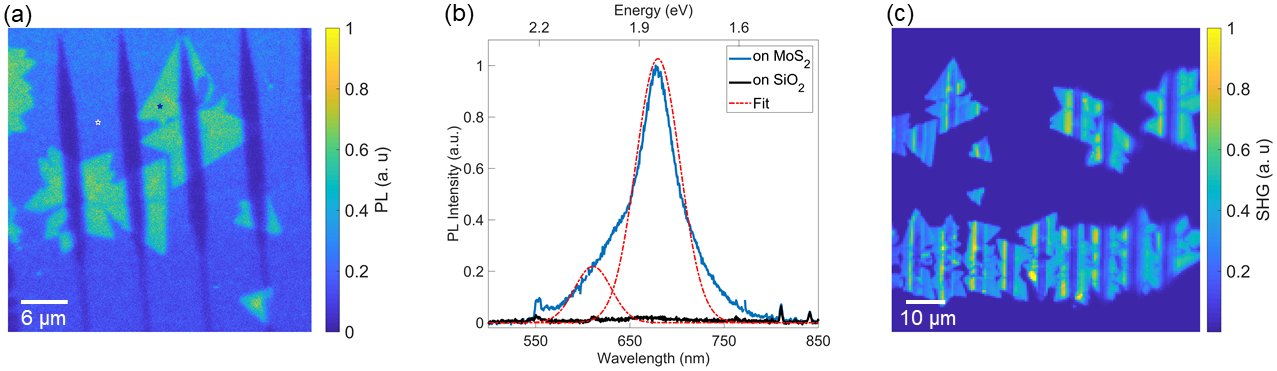}
\caption{Optical characterization of MoS$_{2}$ on waveguide structure. a) Photoluminescence (PL) map of MoS$_{2}$ monolayers on silicon waveguide structure, the spots marked by the ‘*’ denotes the position of spectra measurement. b) PL spectra from the MoS$_{2}$ crystal at the positions marked in Figure 3a; the experimental data is fitted with a double Gaussian fit marked by the dashed red curve. c) Map of second harmonic generation of the MoS$_{2}$ coated waveguide structure.}
\label{fig3}
\end{figure} 

The utilization of TMD-based semiconducting devices in silicon nanostructures not only depends on the TMDs morphology but also on the conservation of its electro-optical properties. We investigated those by mapping photoluminescence (PL) \cite{mak_atomically_2010,splendiani_emerging_2010,eda_photoluminescence_2011,amani_near-unity_2015} and second harmonic generation (SHG) \cite{kumar_second_2013,malard_observation_2013,li_probing_2013}. We choose the combination because SHG is sensitive to structural changes of the monolayer crystals, whereas PL indicates the local electronic properties, dominated, e.g., by doping and defects. Both methods are thus complementary and yield a good overview of the monolayer quality. 

Figure 3a is a PL map of the waveguide structure shown in the optical microscope image in Figure 2a. The TMD crystals covering the SiO$_{2}$ substrate have strong PL. However, the signal is quenched on the Si grating couplers and the Si waveguides. PL spectra could be collected only from MoS$_{2}$ located on SiO$_{2}$. Figure 3b is the room-temperature PL spectrum collected from the positions marked in the PL map. A double Gaussian fitted to the spectrum revealed strong contributions from the A exciton at 1.82 eV and weaker contributions from the B exciton at 2.03 eV. These are consistent with literature values for monolayer MoS$_{2}$ \cite{mak_atomically_2010,splendiani_emerging_2010,eda_photoluminescence_2011,amani_near-unity_2015}. A more detailed analysis is provided in Figure 5 of the supplementary information. Similarly, we map the SHG signal (Figure 3c) after excitation with an 810 nm pulsed laser. A strong SHG response shows that the observed crystals are indeed monolayers. The SHG emission from the grating couplers is slightly suppressed but not as much as the PL emission.
Figure 6c, in the supplementary information provide the power dependency plot of the SHG signal. 
This analysis shows that, although MoS$_{2}$ can be grown with good morphologic properties directly on Si nanostructures, it is not yet suitable for TMD-based semiconducting devices. We hypothesize that the strong PL quenching is related to charge transfer from the substrate to the TMD, as previously observed for 2D material growth on silicon wafer \cite{scheuschner_photoluminescence_2014,man_protecting_2016,ren_-chip_2019} and carrier loss at the TMD-Si interface \cite{bie_mote2-based_2017,li_room-temperature_2017}. Also there is a possibility of un-intentional n-doping of silicon during the lithography assisted fabrication. These mechanisms would not only quench the A- and B-excitons but should also have a detrimental effect on the C-excitons. Besides these factors, silicon has significant absorptive loss at 405 nm \cite{lochner_hybrid_2021}. These explain the observed loss of SHG-efficiency at the 810 nm pump wavelength.

To test this hypothesis and implement a way towards MoS$_{2}$-growth with high optoelectronic quality on Si nanostructures, we created a second batch of samples coated with a passivation layer \cite{ren_-chip_2019}. This layer is for avoiding direct contact between the 2D material and silicon. It is inspired by mechanically exfoliated hBN layers used in a previous study \cite{man_protecting_2016}. The mechanical exfoliation of hBN is not scalable, and CVD growth of hBN requires exceedingly high temperatures ($\approx$ 1000 $\degree$C) \cite{jeong_wafer-scale_2019,glushkov_direct_2021}. Therefore, we coated the second set of nominally identical samples with a protective dielectric layer using atomic layer deposition (ALD) \cite{szeghalmi_tunable_2010,weber_iridium_2011,siefke_materials_2016,shestaeva_mechanical_2017,ghazaryan_structural_2020,lapteva_influence_2021} before the CVD process.

ALD is widely used in the semiconductor industry, and it creates conformal thin films, i.e., it coats steep side walls the same way as planar surfaces \cite{szeghalmi_tunable_2010}. It is an established technique to coat nanostructures with high-quality optical coatings [61,62]. We deposited 5 nm films of Al$_{2}$O$_{3}$ on the Si structures. We chose Al$_{2}$O$_{3}$ as it is one of the widely used substrates for the CVD growth of TMDs \cite{liu_growth_2012,zhang_controlled_2013,ji_unravelling_2015,dumcenco_large-area_2015,chen_oxygen-assisted_2015,aljarb_substrate_2017,yan_chemical_2018}. This thickness, which is larger than the exciton radius provides good electronic separation. But the layer is also thin enough to avoid considerable changes in the optical properties of the waveguide modes, and the overlap of electromagnetic field with the TMD crystals.
\begin{figure}[h!] 
\centering
\includegraphics[width=1\columnwidth]{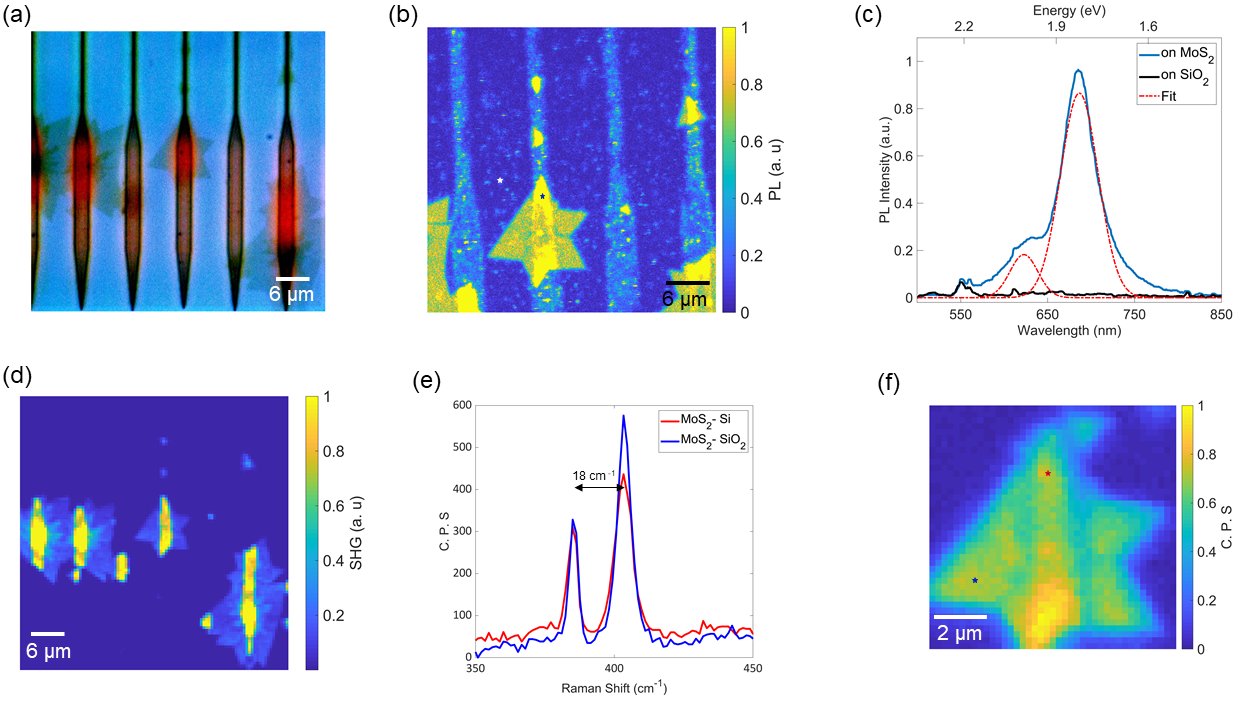}
\caption{Optical characterization of MoS$_{2}$ on waveguide structure with 5 nm Al$_{2}$O$_{3}$. a) Optical microscope image of the sample with a separation layer of 5 nm Al$_{2}$O$_{3}$. b) Photoluminescence (PL) map of MoS$_{2}$ monolayers on silicon waveguide structure, the spots marked by the blue and white stars denote the positions of spectra measurement. c) PL spectra from the MoS$_{2}$ crystal at the positions marked in Figure 4b; the red curve is the double Gaussian fit. d) Map of second harmonic generation of the MoS$_{2}$ coated waveguide structure. e) Raman spectra measured at 532 nm excitation, from two regions; the blue (red) colored spectra at the position labelled by the blue (red) star shown in the Raman map in Figure 4f. The peak positions of the spectra are the same for the MoS$_{2}$ on top of the Si waveguide as well as on SiO$_{2}$. The signal intensity is quantified as photon counts per second (C. P. S. ). f) The map of the E2g peak in the Raman spectra.}
\label{fig4}
\end{figure} 
We grew MoS$_{2}$ on the ALD-coated nanostructure using the CVD process as before. The optical microscope image (Figure 4a) shows triangular TMD crystals. The crystals emit photoluminescence, as shown in Figure 4b. In contrast to the uncoated sample, the PL is equally strong on the silicon waveguides and SiO$_{2}$. This shows that the Al$_{2}$O$_{3}$ barrier prevents charge transfer and recombination at the Si-TMD interface. Figure 4c is the PL spectrum from the regions marked by the stars in the PL map. The Gaussian fitting reveals prominent A excitons at 1.81 eV with an FWHM of 51 nm and weaker B excitons at 1.99 eV with an FWHM of 37 nm as expected from MoS$_{2}$ monolayer crystals of good quality.

We measured the Raman map and Raman spectra to understand the crystal quality further. Raman spectrum is not only sensitive to the number of layers \cite{li_bulk_2012,golovynskyi_exciton_2020} but also to strain \cite{wang_raman_2013,zhou_raman_2014,doratotaj_probing_2016,lee_strain-shear_2017,obrien_raman_2017}. The Raman spectra in Figure 4e were measured from the positions marked in Figure 4f. The spectra showed the characteristic peaks of MoS$_{2}$ around 385.2 cm-1 (E2g) and 403.5 cm-1 (A1g), corresponding to the in-plane vibration of Mo and S atoms and the out-of-plane vibration of the S atoms, respectively \cite{liu_growth_2012,ji_unravelling_2015,dumcenco_large-area_2015}. The separation between the two peaks is 18.3 cm-1, indicating that they are monolayer MoS$_{2}$ crystals. From multiple measurements from different positions on the crystals (Supplementary 9), we conclude that the strain is homogenous over the raised Si waveguides and the lower-lying SiO$_{2}$ substrate. No additional strain is present at the edges as opposed to the case of mechanical transfer.

The SHG map in Figure 4d also proves that the Al$_{2}$O$_{3}$ layer is helpful to retain the optical properties of the MoS$_{2}$. SHG is observable from the entirety of the MoS$_{2}$-crystals. This observation is different from the SHG quenching observed for the uncoated samples. Instead, we find a roughly fourfold SH enhancement on the couplers. While the nature of this enhancement is unclear, we speculate the possibility of local field enhancement on the high-refractive-index Si structures. SH observed from the multilayer crystals at the very right and left of the images suggests that those sections might be AA-stacked multilayers. The Raman spectra (Supplementary 9) also reveals the presence of multilayers on some positions on the waveguide.

\section{Conclusion and Outlook}
We successfully demonstrated the direct CVD growth of conformal monolayer 2D materials on nanostructured silicon on insulator waveguides. We find that the waveguide is functional after the CVD growth process at the intended wavelength (Supplementary 4). The crystals grew around nanometer-sized edges and corners without detectable levels of induced strain. The monolayer crystals exhibit good morphologic properties. Although the interaction of MoS$_{2}$ with silicon reduces its optical properties, a 5 nm coating of Al$_{2}$O$_{3}$ is sufficient to form an electronic barrier between the TMD monolayers and the silicon nanostructures shielding the TMD from the detrimental effect from (doped) silicon. The barrier layer does not influence the optical properties of the waveguides significantly, and it creates an environment to leverage the full potential of the TMD excitons in a scalable manner for integrated optical architectures. We believe that the ramifications may be profound: integrating TMDs directly with silicon waveguides may provide for nanoscale detectors and light sources. This would not only complete the toolbox of integrated wafer-scale components available to silicon integrated architectures, but also allow for the integration of entirely new types of devices such as room-temperature polaritonic lasers \cite{lackner_tunable_2021}, valleytronic devices, or single-photon light sources \cite{peyskens_integration_2019}.

\section{Methods}
\subsection{Waveguide fabrication}
A silicon on insulator wafer was patterned using inductively coupled plasma reactive ion etching (Sentech Instruments Berlin SI 500-C) to create silicon waveguides of 220 nm design height \cite{tuniz_modular_2020}. A positive photoresist mask created by electron beam lithography (Vistec SB 350 OS) is used for the process. The mask alignment is done using an EVG AL-6200. The grating couplers of 50 nm design depth are inscribed by ion beam etching (Oxford plasma technology ionfab 300+LC). An SiO$_{2}$ cladding layer is made by a PECVD process. The cladding layer is then lifted off using a negative photoresist mask. The silicon waveguides discussed here are 10 ${\mu}$m long, 300 nm wide, and have a height of approximately 220 nm above the SiO$_{2}$ cladding. 
\subsection{ALD coating}
The film growth is based on sequential self-limited surface reactions of a metalorganic precursor and an oxidizing or reducing agent. Dielectric thin films are produced using either H$_{2}$O or ozone in thermal processes or O$_{2}$ plasma in plasma enhanced atomic layer deposition (PEALD) at a lower deposition temperature. The optical properties of ALD thin films are comparable to evaporated or sputtered films, and the process optimization allows for the growth of oxides with low optical losses \cite{lapteva_influence_2021}. PEALD is carried out at 100 $\degree$C using an Oxford instruments (Bristol, United Kingdom) OpAL$^{TM}$ ALD reactor. Trimethylaluminum (TMA) and Oxygen (O$_{2}$) plasma have been applied as metal-organic precursor and oxidizing agent, respectively. The optimized PEALD sequence is as follows, 0.03 s TMA pulse, 5 s Argon purge, 5 s O$_{2}$ plasma (300 W, 50 sccm) exposure and finally 2.5 s of Ar purging of chamber leading to a growth per cycle of 1.2 $\text{\r{A}}$/cycle. The film thickness on a reference sample determined by spectroscopic ellipsometry is 6 nm. The typical AFM surface roughness of the coating is about 0.3 nm for a $1{\times}1 {\:\mu}m{^{2}}$ scan area \cite{shestaeva_mechanical_2017,ghazaryan_structural_2020}.
\subsection{CVD growth}
The growth was implemented in a two-zone split tube furnace which allows heating of the precursors individually. The substrate and precursors were placed in an inner quartz tube within the outer tube. The quartz Knudsen cell loaded with Sulphur powder is placed at the centre of the first zone of the furnace. MoO$_{3}$ powder sprinkled on a piece of SiO$_{2}$/Si wafer is placed in the middle of the second zone, and the substrate for growth is placed next to it at the downstream side. Argon gas flow was used to carry the Sulphur to the second zone; where the reaction with MoO$_{3}$ occurs. First zone of the furnace was kept at 200$\degree$ C and the second zone at 730$\degree$ C. The growth process is based on the method explained in reference \cite{george_controlled_2019}. 
\subsection{AFM measurement}
The AFM measurement was performed with a Ntegra (NT-MDT) system in a tapping mode at ambient conditions using n-doped silicon cantilevers (NSG01,NTMDT) with resonant frequencies of 87 - 230 kHz and a typical tip radius of < 6 nm.
\subsection{PL characterization}
The fluorescence mapping was carried out using a confocal fluorescence lifetime imaging microscopy setup (picoquant, microtime 200). The excitation wavelength was 530 nm from a pulsed diode laser at a repetition rate of 80 MHz and the excitation power at the sample was 0.5 mW. The laser is focused on the sample using a 100$\times$ microscope objective (0.85 NA). The PL was collected by the same objective. A 550 nm bandpass filter, and a notch filter that blocks 530 nm laser light is placed before the detector. The detector used for fluorescent mapping is an avalanche photodiode. The microscope objective is connected to a scanning stage; enabling two-dimensional mapping of an area of the sample. The picoquant setup is coupled to a Horiba Jobin Yvon Triax spectrometer, and spectra were measured from different positions of the fluorescent map. The spectrometer utilizes a cooled CCD detector. 
\subsection{SHG measurement}
The second harmonic generation studies were performed using a femtosecond laser emitting pulses with a duration of 100 fs and a repetition rate of 80 MHz (MaiTai, Spectra Physics). The excitation wavelength of 810 nm with vertical polarization was chosen for excitation. The laser power used to pump was 7 mW. A 60$\times$ objective ('Nikon Plan Fluorite Imaging Objective'), with an NA of 0.85 was used to focus the fundamental beam on the sample placed on an x-y piezo controller stage, and the reflected second harmonic signal was collected using the same objective. The collected SH signal was reflected from a dichroic mirror towards the detection path. A short pass filter (600 nm) and a band-pass filter (325-610 nm) were used to allow just the desired wavelength into the CMOS camera (Zyla 4.2 sCMOS). 
\subsection{Raman measurement}
The Raman spectra were measured using a confocal Raman microscope from Renishaw (inVia$^{TM}$ Raman Microscope). The excitation wavelength was 532 nm and a 100$\times$ microscope objective with an NA of 0.85 was used for the measurement.

\begin{acknowledgement}
 A.K. was the principal contributor to the manuscript and responsible for data analysis. F.E. was the overall coordinator of the experiments. A.G., A.Tun., A.Tur., and F.E. developed the concept and contributed to the overall course of the research. A.G. developed the CVD process. Z.G. and E.N. were responsible for the material growth. A.K., Q.N., and S.Shr. were responsible for the PL characterization. Michael Steinert, H.K. and S.Sch. were responsible for the SEM/FIB characterization. P.P. and A.S. were responsible for the ALD process. E.N., N.F., and S.Sch. contributed AFM measurements. F.L. and F.A. are responsible for SHG measurements. T.K. fabricated the waveguides. T.U. and A.K. carried out the Raman measurements. All authors contributed to the manuscript. The authors declare no conflict of interest.
\end{acknowledgement}

\begin{funding}
A.K., H.K., and F.E. are supported by the Federal Ministry of Education and Science of Germany (Grant ID 13XP5053A). A.K., Q.N. and E.N. are supported by the European Social Funds (Grant ID 2019FGR0101), and the Federal state of Thuringia (Grant ID 2018FGR00088). A.G., F.A., F.E., P.P. and A.S. are supported by the Deutsche Forschungsgemeinschaft (DFG, Project-ID 398816777). 
P.P. and A.S. are funded by the Fraunhofer Society (FhG, Attract 066-601020). A. Tun. is supported by an Australian Research Council Discovery Early Career Researcher Award (DE200101041).
\end{funding}

%\bibliographystyle{plane}
%\bibliography{...}

\bibliography{journal-article}

\end{document}